\documentclass[aps,showpacs,floatfix,superscriptaddress,11pt]{revtex4-1}
\usepackage{graphicx}
\usepackage{amssymb}
\usepackage{graphics, color}
\usepackage{amsmath}
\usepackage{overpic}
\usepackage{float}
\usepackage{hyperref}
\usepackage[english]{babel}
\usepackage{hyperref}
\usepackage{amsfonts}
\usepackage[usenames,dvipsnames]{xcolor}
\usepackage{amsmath,amsfonts,amssymb,verbatim,float,mathtools}
\usepackage{physics}

\let\oldhat\hat
\renewcommand{\vec}[1]{\mathbf{#1}}
\renewcommand{\hat}[1]{\oldhat{\mathbf{#1}}}

\begin{document}
\title{Marginal and Irrelevant Disorder in Einstein-Maxwell backgrounds}

\author{Antonio M. Garc\'ia-Garc\'\i a}
\author{Bruno Loureiro}
\affiliation{TCM Group, Cavendish Laboratory, University of Cambridge, JJ Thomson Avenue, Cambridge, CB3 0HE, UK}

\begin{abstract}
We study analytically the effect of a weak random chemical potential of zero average in an Einstein-Maxwell background. For uncorrelated disorder this perturbation is relevant however we show that it can become marginal or even irrelevant by tuning disorder correlations. At zero temperature we find that, to leading order in the disorder strength, the correction to the conductivity for irrelevant perturbations vanishes. In the marginal case, in order to renormalize a logarithmic divergence, we carry out a resummation of the perturbative expansion of the metric that leads to a Lifshitz-like geometry in the infrared. Disorder in this case also induces a positive correction to the conductivity. At finite temperature the black hole acquires an effective charge and the thermal conductivity has the expected Drude peak that signals the breaking of translational invariance. However the electric conductivity is not affected by the random chemical potential to leading order in the disorder strength.  
\end{abstract}
\date{\today}

\maketitle
\section{Introduction}
Disorder plays an important role on the transport properties of interacting electrons in solids. A small amount of disorder in systems with translational symmetry makes the direct current conductivity finite. Similarly, disorder slows down the classically diffusive dynamics of electrons in solids at finite temperature. In real materials disorder is typically introduced by chemical doping which in some cases obscures its effect: the conductivity may increase because the slow down of the motion caused by disorder is counterbalanced by the addition of new carriers.

By contrast, in the limit of vanishing temperature and interactions quantum coherence phenomena enhance dramatically the effect of disorder.   
According to the one parameter scaling theory of localization \cite{Abrahams1979}, classical diffusion in two and lower dimensions is completely arrested for any disorder and sufficiently long times. This quantum coherence phenomenon, usually referred to as Anderson localization \cite{Anderson1958}, also  occurs in higher dimensions \cite{Frohlich1983} for sufficiently strong disorder. The metal-insulator transition at finite disorder is characterized by universal critical exponents \cite{Vollhardt1980,Vollhardt1982,Garcia2008}. Overwhelming numerical \cite{Rodriguez2011, Slevin2014, Garcia2007}, analytical \cite{Frohlich1983,Abou-Chacra1973}, and more recently experimental \cite{Roati2008,Billy2008} evidence, from cold atom physics where interactions can be tuned to be negligible, have all but confirmed the predictions of the scaling theory of localization in the non-interacting limit. However in real materials there are always interactions that may potentially weaken or completely destroy Anderson localization. For two spatial dimensions a diagrammatic resummation showed \cite{Gorkov1979} that constructive interference between clockwise and counter clockwise loops, the so called weak-localization corrections, induces a logarithmic increase of the resistivity for sufficiently low temperatures \cite{Gorkov1979}. Interestingly the effect of weak interactions in a weakly disordered potential, neglecting coherence effects, causes a similar log increases though with a different prefactor \cite{Altshuler1980}. Therefore, in this limit at least, it seems that interactions do not destroy weak-localization which is in full agreement with experimental results. More recently, interest has shifted to the stability of full Anderson localization in the presence of interactions. 
Qualitative calculations \cite{Basko2006a,Basko2006} in the physics literature and more rigorous, but restricted to mean-field interactions, mathematical results \cite{Wang2008} agree that Anderson localization for sufficiently strong disorder still persists in the presence of weak interactions. 
This novel state of quantum matter, usually referred to as many body localized, is strictly an insulator since the conductivity vanishes in the limit of zero frequency and temperature. However it has still some distinctive dynamical properties like logarithmic, instead of linear, growth \cite{Bardarson2012} of the entanglement entropy after a quench, vanishing of the ac conductivity as a power law, $\sigma \propto \omega^\alpha$ with $ 0<\alpha \leq 2$ without logarithmic corrections \cite{Gopalakrishnan2015} or glassy features like the possibility of slow logarithmic diffusion \cite{Wang2008}. A detailed understanding of the interplay between disorder and interactions is seriously hampered by computational limitations and the lack of analytical tools to tackle strong interactions. 

Holographic dualities \cite{Maldacena1997}, that propose that certain strongly coupled field theories in $d$ dimensions are dual to classical theories of gravity in $d+1$ dimensions, offer a promising framework to tackle this problem. 

Indeed there are already several studies of the role of disorder in a strongly coupled field theory with a gravity-dual. Originally disorder was introduced \cite{Hartnoll2008b,Fujita2008,Hartnoll2012} as a deformation of the boundary field theory that coupled the random potential to an operator of the conformal field theory. The addition of this perturbation breaks translational invariance so effectively the role of disorder was to induce momentum relaxation which alters substantially the transport properties of the dual field theory. 

In the context of holographic superconductors the effect of a random chemical potential has been studied numerically  but only in the probe limit where disorder does not backreact in the metric \cite{Arean2014,Zeng2013}. Disorder has also been considered in hyperscaling violating backgrounds, also in the probe limit \cite{Lucas2014,Lucas2015,Lucas2015a}.
Backreaction effects of a weak but marginally disordered scalar at zero temperature leads to logarithmic divergences in the infrared that suggest an instability of the perturbation theory \cite{Adams2011,Adams2014}. However it was later proposed \cite{Hartnoll2014} that that these divergences were an artefact of the perturbation theory in non-linear problems that could be cured by the Poincar\'e-Lindstedt method. The resulting metric in the infrared, after an effective resummation of logarithmic corrections, becomes Lifshitz-like with a dynamical critical exponent that depends on the strength of disorder. In the infinite temperature case \cite{Hartnoll2015,Hartnoll2015a} it seems that the presence of a horizon prevents any Lifshitz scaling in the infrared. Numerical simulations for stronger disorder \cite{Hartnoll2014,Hartnoll2015a}, still for a disordered scalar at zero and finite temperature, have not shown any qualitative change to these results. 

In the context of Einstein-Maxwell theories it has been recently proposed \cite{Donos2014b,Donos2015} a general expression for the averaged conductivity in gravity-duals, modified by disorder or any other source of inhomogeneity, in terms of the solution of the Einstein equations for the metric. The study of these solutions has just started: the effect of weak disorder in the Einstein-Maxwell theory induced by a random chemical potential including backreaction effects, recently studied in Ref.\cite{OKeeffe2015}, reveals surprising features like a conductivity that increases with disorder. We note that the disorder investigated in Ref.\cite{OKeeffe2015} is a relevant perturbation that leads to linear, instead of logarithmic, divergences in the metric. Although the Poincar\'e-Lindstedt method is technically applicable in this case it is less clear that these divergences are really an artefact of the perturbation theory. 

Here we revisit this problem by studying an Einstein-Maxwell background with a random, but in general correlated, chemical potential of zero average at zero and finite temperature. By modifying the correlations of the disordered chemical potential we tune the conformal dimension of the gauge field so that we can also investigate irrelevant and marginal
perturbations. In the limit of zero temperature we have found that, to leading order, irrelevant perturbations do not modify the conductivity in the limit of zero temperature. By contrast, for marginal disorder the corrections to the conductivity are positive. In this case the metric develops perturbative logarithmic singularities in the infrared that can be resummed by using the Poincar\'e-Lindstedt method \cite{Hartnoll2014}. The resulting geometry is Lifshitz-like in the infrared with a dynamical critical exponent that depends on the disorder strength. In the finite temperature case the effect of perturbative disorder is weaker. The electrical conductivity does not get corrections in the disorder strength to leading order though the black-hole becomes charged even in this limit. 

We start by introducing the Einstein-Maxwell theory with a random but correlated chemical potential. 

\section{Correlated Disorder in the Einstein-Maxwell background}
\label{model}
We investigate the interplay of disorder and interactions in field theories with a gravity dual. 
For that purpose we study an asymptotic anti-de Sitter ($AdS$) Einstein-Maxwell theory in $d+1=4$ space-time dimensions with a random chemical potential given by the action 
\begin{equation}
\label{action}
S=\int d^{4}x\sqrt{-g} \left[ R + 6 -\frac{1}{4} F^2\right],
\end{equation}
\noindent where $F$ is the Maxwell tensor, $R$ the scalar curvature. For convenience we have set $l_{\text{AdS}_4}=2\kappa^2_4 = 1$. We choose to work in Fefferman-Graham coordinates $\dd s^2 = \frac{1}{z^2} \left(\dd z^2 + g_{\mu\nu}(x^{\mu},z)\dd x^{\mu}\dd x^{\nu} \right)$ which we suppose are globally defined. Here $z=0$ is the AdS boundary, with coordinates $x^{\mu} = (t,x,y)$. The equations of motion are given by
\begin{subequations}
\label{eemaxwell}
\begin{align}
R_{ab} + 3 g_{ab} = \frac{1}{4}F_{\phantom{c}a}^{c}F_{bc} - \frac{1}{8} g_{ab} F^2 \label{emaxwelleq}, \\
\partial_{a}\left( \sqrt{-g}F^{ab} \right) = 0 \label{maxwelleq}.
\end{align}
\end{subequations}
We are only interested in spatially inhomogeneous solutions of the equations above, for which neither the metric components nor the gauge field depend on time. Therefore the $U(1)$ gauge field $A = a_t(\vec{x},z) \dd t$, which we assume is the only non-zero component, and the metric $g_{\mu\nu}(\vec{x},z)$ depends explicitly on the bulk ($z$) and the boundary spatial coordinates ($\vec{x}$). 
This system of equations support both zero and finite temperature solutions, which are specified by the infrared (IR) and ultraviolet (UV) boundary conditions. For the zero temperature case, we require all metric components and the gauge field to be regular at the Poincar\'e horizon $z\to\infty$. For the finite temperature case, we require the existence of a horizon, \emph{i.e.} a point $z_0\in (0,\infty)$ such that $g_{tt}(\vec{x},z) \sim \gamma_{tt}(\vec{x})(z-z_0)+O\left((z-z_0)^2\right)$ and $g_{zz}(\vec{x},z) \sim \frac{\gamma_{zz}(\vec{x})}{z-z_0}+O\left(1\right)$ for $|z -z_0| \ll 1$, with all other components $g_{ij}$ regular.
Similarly, close to the boundary we impose,   
\begin{subequations}
\label{boundaryconditions}
\begin{align}
\lim\limits_{z\to0} g_{ab} \dd x^a \dd x^b&= \frac{1}{z^2} \left( \dd z^2-\dd t^2+\dd x^2+\dd y^2  \right) , \label{bcmetriccharged}\\
\lim\limits_{z\to 0} a_t(\vec{x},z)&= \mu(\vec{x}) \label{bcvectorpotential}.
\end{align}
\end{subequations}
According to the holographic dictionary the bulk action (\ref{action}), with the above boundary conditions, is dual to a $d=3$ conformal field theory (CFT) at finite chemical potential $\mu(\vec{x}) = \lim\limits_{z\to 0}a_t(x,z)$. 
Disorder is introduced, in one or both boundary directions, through a random chemical potential in the boundary $\mu(\vec{x})$. Next we give a detailed account of the properties of this random chemical potential so that we can use it to model irrelevant and marginal perturbations in the dual field theory.

\subsection{Correlated Disorder and relevance of perturbations}
\label{impdisorder}
We introduce disorder in the holographic setting by imposing that the chemical potential $\mu(\vec{x})$ is a stochastic field depending on the spacelike boundary coordinates. This random boundary condition promotes the vector potential $A=a_t(z,\vec{x})\dd t$ and the metric components to stochastic processes indexed by $\vec{x}$. Similarly the equations of motion \eqref{emaxwelleq} and \eqref{maxwelleq} becomes stochastic equations.

We specify the distribution of $\mu(\vec{x})$ by a spectral decomposition
\begin{equation}
\label{bcvp}
\mu(\vec{x}) = \bar{V} \int_{\mathbb{R}^n} \frac{\dd[n] k}{(2\pi)^n} ~e^{i \vec{k}\cdot \vec{x}} \mu_\vec{k},
\end{equation}
\noindent where $n=1$ if disorder is only in one direction or $n=2$ if disorder is in both directions and the parameter $\bar{V}$ measures the amplitude of the source ($\mu\sim O(\bar{V})$). Further, we assume $\mu_\vec{k}$ is a spectral stochastic process taking values in a Gaussian distribution 
of zero average $\mathbb{E}[\mu_\vec{k}]=0$ and variance $\sigma_\vec{k}^2$ where $\mathbb{E}[\dots]$ denotes the average with respect to the probability distribution. 
From now on we will restrict ourselves to isotropic disorder $\mu_\vec{k} = \mu_k$ so that Eq. \eqref{bcvp} can be  written effectively as a one dimensional integral $\int_{\mathbb{R}^{n}} \dd[n]k = \text{Vol}(\mathbb{S}^{n})  \int_{0^+}^{\infty}\dd k~ k^{n-1}$.
We stress that even though $\mu_k$ is Gaussian, this does not imply $\mu(\vec{x})$ is Gaussian itself unless $\sigma_{k}^2$ is $k$ independent. It was shown recently \cite{OKeeffe2015} that precisely in this case even a weakly disordered chemical potential $\bar{V} \ll 1$ induces a relevant perturbation in the geometry which casts some doubts on the reliability of the perturbation theory. 

Interestingly the relevance, or not, of the perturbation depends on the disorder correlations as the mass dimension of $\bar{V}$ is controlled by the mass dimension of $\sigma_k^2$. Specifically, we have $[\mu] = [\bar{V}]+n+[\mu_k] = [\bar{V}] + \frac{1}{2}(n+[\sigma_{k}^2])$. Therefore introducing powers of $k$ in $\sigma_k$ makes disorder more and more irrelevant. For instance assuming $\sigma_k^2 \propto k^s$ 
\begin{equation}
\label{power}
[\bar{V}]= 1- \frac{n+s}{2}.
\end{equation}
Therefore disorder is relevant ($\bar{V}>0$) if $n+s < 2$, marginal ($\bar{V}=0$) if $n+s = 2$ and irrelevant  ($\bar{V}<0$) if $n+s > 2$. 
In the following we will restrict to marginal and irrelevant perturbations by employing correlated potentials such that $n+s \geq 2$. Surprisingly, we shall see that this \emph{a priori} naive power counting actually determines the perturbative flow of the renormalization group (RG) in the Einstein-Maxwell system.
Finally we note that for a fixed $s$, increasing the number of dimensions in which we introduce disorder makes disorder less relevant. The fact that translation invariance is left unbroken in a bulk direction constrains the dynamics of the fields to the orthogonal directions. It is therefore no surprise that disorder is more relevant in this case. Indeed it is a well known result in condensed matter systems that disorder is more relevant in lower dimensional systems \cite{Abrahams1979}.

\subsection{Explicit implementation of disorder}
\label{iranduv}
We have now all the ingredients to define the correlated disordered potential to be employed in the rest of the paper.  
For most of the analytical calculations we shall employ Eq.\eqref{bcvp} assuming isotropic disorder \footnote{When working in higher dimensions, we denote $k=|\vec{k}|$. In one dimension we explicitely include the modulus to avoid ambiguities.} and a Gaussian $\mu_k$ with zero average $\mathbb{E}[\mu_k]=0$ and variance
\begin{equation}\label{variance}
\sigma_k^2 = 2^{s+1} k^s e^{-2ka}.
\end{equation}
Note that the exponential factor assures convergence of the boundary deformation by smoothly suppressing high momenta modes. This introduces a UV length scale $a =1/k_0$, necessary to cure divergences for irrelevant perturbations, which can be interpreted as a lattice constant that effectively suppresses modes with wavelength smaller than the lattice spacing. We stress that since we are interested on averaged quantities that are computed analytically it is not necessary an explicit expression for $\mu(\vec x)$.

However in the finite temperature case we shall find more convenient at times to employ the following explicit representation of the random chemical potential commonly used in the holography literature \cite{Zeng2013, Arean2014, Arean2014a, Arean2015, Hartnoll2014, Hartnoll2015, Hartnoll2015a, OKeeffe2015}
\begin{equation}
\label{discrep}
\mu(\vec{x}) =\bar{V} \sum\limits_{\{ m_i \} =1}^{N-1} A_{\{m_i\}}\prod\limits_{i=1}^{n} \cos(k_{m_i}x^i + \gamma_m).
\end{equation}
\noindent where $A_{\{m_i\}} = \bar{V}(\sqrt{\Delta k \sigma_{\{m_i\}}})^n$ with  $\Delta k = k_0/N$ and $k_{m_i} = m_i \Delta k$. Here $\gamma_m\in [0,2\pi)$ are i.i.d. random variables. Further, we define $\Delta k = k_0/N$ and $k_{m_i} = m_i \Delta k$. Averages $\mathbb{E}[\cdots]$  in this representation are taken with respect to the i.i.d distribution of phases $\gamma_m$, the variance is given by Eq. \eqref{variance} though the UV cutoff $k_0=1/a$ is sharp and applied directly to the sum. Note that in this representation there is also a natural IR scale $k_* = 1/L = 1/{Na}$ which is only taken to zero in the averaging procedure.

Both the discrete and the continuous representations are equivalent in the limit $a\to 0$ and $L\to\infty$. For finite values of the cutoffs we still expect qualitatively similar results.

\section{Random chemical potential at zero temperature}
In this section we study the $d+1=4$ Einstein-Maxwell action at zero temperature in the presence of a weak and correlated random chemical potential. We investigate the cases of disorder acting in one and two boundary space dimensions. Although both cases are quantitatively different, they have a similar IR behaviour as long as correlations are chosen so that disorder is marginal. For marginal disorder we find logarithmic IR divergences in the metric that can be resumed by the Poincar\'e-Lindstedt method leading to a Lifshitz-like metric. We proceed with the calculation of the DC conductivity for both irrelevant and marginal disorder. We find the perturbative correction vanishes for irrelevant disorder and is positive for marginal disorder. The divergence of the marginal flow signals an instability of the system towards, possibly to a charged ground state at finite temperature.

\subsection{Metric corrections for disorder in one dimension}
Consider the action \eqref{action} with boundary conditions \eqref{boundaryconditions} at zero temperature. We fix coordinates $x^{\mu}=(t,x,y)$ in the boundary and restrict disorder to act only in the $x$ direction. Following the discussion in section \ref{impdisorder}, we introduce disorder by requiring $\mu(x)$ to be a homogeneous random field with spectral decomposition
\begin{equation*}
\mu(x) = \bar{V} \int_{\mathbb{R}} \frac{\dd k}{2\pi} ~e^{i k x} \mu_k,
\end{equation*}
\noindent where $\mu_k$ is a gaussian spectral process with zero mean. Finding exact solutions of the system \eqref{eemaxwell} is a hard task so we restrict ourselves to a perturbative analysis in disorder strength $\bar{V}$. Acording to Eq.\eqref{power}, for $\sigma_k = 1$ we have $[\bar{V}] = 1/2 > 0$ and therefore disorder is relevant in this case. A perturbative analysis is therefore inadequate, as disorder can drive the theory to a new fixed point far from $AdS_4$. This lead us to consider correlated disorder with $\sigma_k = 2^{s+1}|k|^{s}e^{-2|k|a}$. It is easy to see that by choosing $s = 1$ disorder will be marginal. Therefore we might be able to find new disordered fixed points close to $AdS_4$ by adapting the analysis of Ref. \cite{Hartnoll2014} for a scalar coupled to gravity to the case of Einstein-Maxwell theory.  

To set up the perturbation theory, we write the most general static line element in Fefferman-Graham coordinates compatible with our boundary conditions
\begin{equation*}
ds^2 = \frac{1}{z^2}\left[-A(x,z)\dd t^2 + \dd z^2 + B(x,z)\dd x^2 + D(x,z)\dd y^2 \right],
\end{equation*}
\noindent and proceed with a perturbative expansion in $\bar{V}\ll1$
\begin{subequations}
\begin{align*}
A(x,z) &= 1 + \bar{V}^{2} \alpha(x,z) + O(\bar{V}^2), & B(x,z) &= 1 + \bar{V}^{2} \beta(x,z) + O(\bar{V}^2), \\ 
D(x,z) &= 1 + \bar{V}^{2} \delta(x,z) + O(\bar{V}^2), & a_t(x,z) &= \bar{V} \varphi(x,z) + O(\bar{V}^3),
\end{align*}
\end{subequations}
\noindent where all $\alpha,\beta,\delta,\varphi$ have been lifted to stationary stochastic processes via the boundary conditions and Einstein's Equations. Note that to order $\bar{V}^0$ the background is pure $AdS_4$. To order $\bar{V}^1$, Maxwell's Equation \eqref{maxwelleq} is a Laplace equation
\begin{equation*}
\partial_z^2\varphi + \partial_x^2 \varphi = 0,
\end{equation*}
\noindent which can be solved by decomposing $\varphi(x) = \int \frac{\dd k}{2\pi}~ e^{i k x}\varphi_k(z)$ and imposing the boundary conditions \eqref{bcvp} together with regularity at $z\to\infty$:
\begin{equation}
\label{solvp}
\varphi(x,z) = \int \frac{\dd k}{2\pi}~e^{-|k|z+ikx}\mu_k .
\end{equation}

We now need to insert this into the $O(\bar{V}^2)$ Einstein's Equations, that can be reorganized to give:
\begin{subequations}
\label{eom}
\begin{align}
\partial_z \left[z^{-2}\partial_z\left(\alpha+\delta\right) \right] &= \frac{1}{2}\left[(\partial_z\varphi)^2 -(\partial_x\varphi)^2 \right] \label{eq1},\\
z^2 \partial_z \left(z^{-1}\partial_z\beta\right) &= -\partial_z\left(\alpha+\delta \right)  \label{eq2},\\
\partial_z\partial_x(\alpha+\delta) &= z^2 \partial_z\varphi\partial_x\varphi  \label{eq3},\\
2z^3 \partial_z \left[z^{-2}\partial_z\left(\alpha-\delta\right) \right] + 2z \partial_x^2\left(\alpha-\delta\right) &= 2z^3\left[(\partial_z\varphi)^2 +(\partial_x\varphi)^2 \right].  \label{eq4}
\end{align}
\end{subequations}

In practice this can be solved explicitly by inserting Eq. \eqref{solvp} in the right hand side of the above equation, developing $\alpha,\beta,\gamma$ in harmonics and integrating the resulting EOM's. However, since we are not interested in the specific realizations of the random geometry but rather in the possible IR averaged fixed points, we take the average of the above equations:
\begin{subequations}
\begin{align}
\label{averages}
\mathbb{E}[(\partial_z\varphi)^2 -(\partial_x\varphi)^2]&= 0, \\ 
\mathbb{E}[(\partial_z\varphi)^2 +(\partial_x\varphi)^2] &= 
\int_{0^+}^{\infty} \frac{\dd k}{2\pi}~  2^{s+2} k^{s+2} e^{-2 k(z+a)}= \frac{\Gamma(s+3)}{4\pi(z+a)^{s+3}}\label{avemt}.
\end{align}
\end{subequations}
\noindent where we assumed $s>-3$ \footnote{We are not interested in the range $s<0$ since we already know disorder is relevant in this case a perturbative approach is not adequate.}. From the above it is clear that the solutions of equations \eqref{eq1} and \eqref{eq2} are regular for all $z\ge0$ while solutions of \eqref{eq4} can develop divergences depending of the value of $s$. Explicitly we have:
\begin{subequations}
\begin{align}
\mathbb{E}[\alpha+\delta] &= \eta, \label{alphadelta}\\
\mathbb{E}[\beta] &= \eta, \label{beta}\\
\mathbb{E}[\alpha-\delta] &= -\frac{\Gamma(s+3)}{4\pi(s+2)}\int \frac{\dd z}{(z+a)^{s}}\propto\left\{\begin{matrix}
\log (z+a) & \text{ for } s=1, \\ 
(z+a)^{1-s} & \text{ for } s\neq 1.
\end{matrix}\right. \label{analysisdisorder}
\end{align}
\end{subequations}
\noindent where we have imposed regularity at $z\to\infty$ and the boundary conditions $\mathbb{E}[(\alpha-\delta)(x,0)] = \text{const}$. Note that this result reproduces exactly what we naively expect from the power counting analysis: for $s<1$, disorder is relevant and therefore the perturbation scheme breaks down with the appearance of power law divergences in the deep IR $z\to\infty$. For $s>1$, disorder is irrelevant, and indeed the background flows to pure $AdS_4$ in the IR. For $s=1$ disorder is marginally relevant, as signaled by a log divergence as we flow towards the IR. This log behavior was first observed in  \cite{Adams2014} and later reproduced in \cite{Hartnoll2014, Hartnoll2015} in the case of a disordered scalar. Our analysis for the charged case suggests that the log divergences for marginal deformations are a quite general feature of holographic disorder.  

\subsubsection{Resummation of the metric for marginal disorder}
\label{secresum}
In a perturbative RG analysis, one is interested in how the deformation of a given action can change the IR behaviour of the theory. Divergences signal an instability of the flow towards new fixed points. In particular, logarithmic divergences are usually associated with marginal deformations which can sometimes be resummed, to all orders, to give the explicit IR effective action \cite{Cardy1996}. A similar procedure to resum log divergences in Holography was first proposed by Hartnoll and Santos in \cite{Hartnoll2014}. As was mentioned in the introduction, the upshot is that log divergences in holography are associated with IR geometries that can be characterized by their scaling properties. In the case of scalar deformations, they found an emergent Lifshitz scaling with dynamical critical exponent $\bar{z}(\bar{V})$ which is an increasing function of disorder. 

The general idea is to modify the metric ansatz by including a function that regularize the divergences order by order in perturbation theory, similar to the Poincar\'e-Lindsteadt method used in the study of non-linear oscillators. Our ansatz is
\begin{equation}
\label{ansatzresum}
ds^2 = \frac{1}{z^2}\left[-\frac{A(x,z)}{F_1(z)^{p(\bar{V})}}\dd t^2 + \dd z^2 + B(x,z)\dd x^2+ \frac{D(x,z)}{F_2(z)^{q(\bar{V})}}\dd y^2\right],
\end{equation}
\noindent and consists of corrections only to the IR diverging components of the metric. Since divergences appear in the second order of perturbation theory, we can expand $p(\bar{V}) = p_2 \bar{V}^2+O(\bar{V}^4)$ and $q(\bar{V}) = q_2 \bar{V}^2+O(\bar{V}^4)$ and require $\lim\limits_{z\to 0}F_{1,2} = 1$ in order to preserve the UV physics. The equations of motion \eqref{eom} now read:
\begin{subequations}
\begin{align*}
\partial_z \left[z^{-2}\partial_z\left(\alpha+\delta-\log F_1^{p_2}F_2^{q_2}\right) \right] &= \frac{1}{2}\left[(\partial_z\varphi)^2 -(\partial_x\varphi)^2 \right],\\
z^2 \partial_z \left(z^{-1}\partial_z\beta\right) &= -\partial_z\left(\alpha+\delta-\log F_1^{p_2}F_2^{q_2} \right) ,\\
\partial_z\partial_x(\alpha+\delta) &= z^2 \partial_z\varphi\partial_x\varphi ,\\
z^2 \partial_z \left[z^{-2}\partial_z\left(\alpha-\delta-\log \frac{F_1^{p_2}}{F_2^{q_2}}\right) \right] +  \partial_x^2\left(\alpha-\delta\right) &= z^2\left[(\partial_z\varphi)^2 +(\partial_x\varphi)^2 \right]. 
\end{align*}
\end{subequations}
From the above it is clear that choosing $F_1 = F_2 = F$ and tuning $p = 1/2 = -q$ leaves $\alpha + \delta$ and $\beta$ unchanged while shifts $\alpha-\delta \to \alpha-\delta - \log F(z)$ by the log of an arbitrary function $F(z)$. Any choice of $F(z)$ satisfying the constraint $F(0) = 1$ and such that $F(z)\sim z$ as $z\to\infty$ will regularize the IR log divergence previously found (\emph{e.g.} $F(z) = 1+ (z/a)^2$). Up to a rescaling of the coordinates by a constant, the averaged IR metric can then be written as:
\begin{equation*}
\mathbb{E}[ds^2_{IR}] \sim -\frac{\dd t^2}{z^{2b_1}} + \frac{\dd z^2 + \dd x^2}{z^2}+\frac{\dd y^2}{z^{2b_2}},
\end{equation*}
\noindent for $b_1 = 1+\bar{V}^2/2 + O(\bar{V})^4$ and $b_2 = 1-\bar{V}^2/2 + O(\bar{V})^4$. The emergent IR metric has an anisotropic scaling symmetry in the bulk directions. This should not be a surprise since isotropy is broken by disorder. Next we show isotropy is recovered by considering disorder in both boundary space directions.

\subsection{Metric corrections for disorder in two dimensions}
\label{sectiondisorder2d}
In the previous section, we found that working with the averaged geometry is enough to determine the instability of the RG flow. The way the metric diverges is intimately connected to the emergent scaling behaviour of the IR disordered fixed points. In this section we show that a similar log divergence emerges when disorder is considered in all space boundary directions. The advantage is that isotropy is recovered, making easier to generalize to higher dimensions and finite temperature. As expected, the resulting metric has an emergent Lifshitz scaling in the IR.

The framework used above can be generalized to include disorder in both bulk directions $(x,y)$ with the changes:
\begin{align}
k \to\vec{k} = (k_x, k_y), & & x\to\vec{x}=(x,y), & & \int \dd k \to \int \dd[2] k. \nonumber
\end{align}
The power counting \eqref{power} now give us $[\bar{V}] = -s/2$, and disorder is marginal for $s=0$. This is not surprising, since by performing the above changes in Eq. \eqref{avemt}, it is clear that the double integral contributes with an additional power of $k$.

The only difference in the averaged equation of motion is the appearance of non-trivial $y$ dependence in the gauge field $A$ components. They can be conveniently rearranged as \footnote{To avoid charging the notation, we conveniently denote $\mathbb{E}[\alpha(x,z)] = \alpha(z)$, etc.}:
\begin{subequations}
\label{eom2}
\begin{align}
4\partial_z \left[z^{-2}\partial_z\alpha \right] &=\mathbb{E}\left[3(\partial_z\varphi)^2 +(\nabla\varphi)^2 \right]=\frac{3}{2\pi}(z+a)^{-4} \label{eq42d}, \\
2\partial_z \left[z^{-2}\partial_z\left(\beta+\delta\right) \right] &=-\mathbb{E}\left[(\partial_z\varphi)^2 +(\nabla\varphi)^2 \right]=-\frac{3}{4\pi}(z+a)^{-4} \label{eq32d}, \\
\partial_z \left[z^{-2}\partial_z\left(\beta-\delta\right) \right] &=\mathbb{E}\left[(\partial_x\varphi)^2 -(\partial_y\varphi)^2 \right]=0 \label{eq22d}, \\
2\partial_z \left[z^{-2}\partial_z\left(\alpha+\beta+\delta\right) \right] &=z~ \mathbb{E}\left[(\partial_z\varphi)^2 -(\nabla\varphi)^2 \right]=0 \label{eq12d},
\end{align}
\end{subequations}
\noindent where we introduced the bulk gradient $\nabla = (\partial_x,\partial_y)$. As advertised, now disorder does not break isotropy in the bulk, and this is reflected in the equations of motion \eqref{eq32d}. In the one dimensional case $\mathbb{E}[\partial_y\varphi]=0$ and the average in the right hand side do not vanish. The equations above can be easily solved to give:
\begin{align*}
\alpha = -(\beta+\delta)= -\frac{1}{8\pi}\log (z+a),
\end{align*}
\noindent which is in agreement with marginally relevant deformations. In analogy with the one dimensional case, it is again possible to resum these logarithmic corrections. Up to a coordinate redefinition, the IR geometry will take the form:
\begin{equation*}
 \mathbb{E}[ds^2_{IR}] \sim -\frac{\dd t^2}{z^{2\bar{z}}} + \frac{\dd z^2}{z^2} + \frac{\dd x^2 + \dd y^2}{z^{2}}.
\end{equation*}
\noindent with $\bar{z}= 1 + \bar{V}^2 + O(\bar{V}^2)$. This IR fixed point corresponds to a quantum field theory with \emph{Lifshitz} scaling, since it is invariant under $(t,x,y)\to (\lambda^{\bar{z}}t, \lambda x, \lambda y)$. The emergence of Lifshitz scaling in the context of disordered holography was first observed in Ref. \cite{Hartnoll2014}. It is an interesting fact that Lifshitz-like scaling emerges in different dimensions and for different random sources. This suggests that Lifshitz geometries in the IR are a robust feature of marginal disorder in holography.

\subsection{Conductivity of the dual field theory}
\label{sectioncond}
We now turn our attention to how disorder affects the transport of the dual theory.  In condensed matter, the effect of weak disorder in a metal is to decrease the conductivity \cite{Kramer1993}. This is the first sign that for strong enough disorder the system undergoes a metal-insulator transition. According to the \emph{scaling theory of localization} \cite{Abrahams1979, Anderson1980} (or \cite{Kramer1993} for a review) the knowledge of the scaling of the conductance with the system size allows to derive a real space RG equation and eventually to establish the existence of the metal-insulator transition. 

In holography, it was established that a range of theories in both zero and finite temperature have a finite and constant incoherent contribution to the conductivity in addition to the usual coherent contribution coming from a finite charge density \cite{Davison2015a}. In particular this contribution is also present at zero temperature and charge density \cite{Policastro2002, Iqbal2009}. Our aim is to understand how disorder affects this contribution. For simplicity, we work with disorder in one dimension and compute the DC conductivity in this direction. We will show that an irrelevant disordered chemical potential does not contribute to the conductivity, while a marginal deformation has the effect of increasing it. 
In both cases disorder does not suppress the incoherent contribution to the conductivity. It is an open question why those degrees of freedom seem to be protected from relaxation. In principle this is different from the behaviour expected in condensed matter systems where disorder always suppresses the conductivity. We note that a direct comparison is difficult as our perturbation may also induce a net increase of carriers that enhance the conductivity. 

Computing transport coefficients in inhomogeneous backgrounds is an involved task. Since we are only interested in the DC conductivity, we are going to take a shortcut first proposed by Donos and Gauntlett in Ref. \cite{Donos2014b} which consist in applying a constant electric field $E^x\equiv E$ in the disordered direction at the dual boundary theory. In the bulk, this is implemented by a fluctuation in the vector potential that solves the time dependence of the Maxwell's Equations,
\begin{align*}
\delta A = (a_x(x,z)-Et)\dd x.
\end{align*}
This fluctuation generates a non-trivial boundary current obtained via the usual holographic dictionary $j^x = \lim\limits_{z\to 0}\partial_z a_x$. The conductivity is then defined as
\begin{align}
\label{conductivity}
\sigma = \left.\frac{\mathbb{E}[j^x]}{E}\right|_{z=0}.
\end{align}
The fluctuation above also couples to the metric via the Einstein's Equations, and consistency require turning on metric fluctuations. In a radial gauge $h_{za} = 0$ for $a\in\{z,t,x,y\}$, the Einstein's Equations decouple in two sectors, and is sufficient to consider only the metric fluctuation $h_{tx}$.

As in the previous section, we proceed with a perturbative analysis. Inspection of the equations of motion to first order in $E$ and second order in $\bar{V}$ requires 
\begin{align*}
a_x(x,z)=a_x^{(0)}(x,z)+\bar{V}^2 a_x^{(2)}(x,z) +O(\bar{V}^4),&& h_{tx}(x,z)=\bar{V} h_{tx}^{(1)}+O(\bar{V}^3).
\end{align*}
Note in particular that we need to expand the fluctuation $a_x$ to $O(\bar{V}^2)$ in order to respect the holographic dictionary and match the boundary current $j^x = \lim\limits_{z\to 0}\partial_z a_x$ to the bulk current $\sqrt{-g}F^{xz}$. To compute conductivity \eqref{conductivity} we need to solve the Einstein-Maxwell order by order for $\{a_x^{(0)},a_x^{(2)},h_{tx}^{(1)}\}$ and take the relevant average over disorder. As boundary conditions for the fluctuations, we require $\delta A_x$ to be ingoing and the dual field theory Minkowski metric to be fixed, or in other words $\lim\limits_{z\to0}z^2 h_{tx} = 0$\footnote{We are grateful to Andrew Lucas for pointing this out.}.

To order $\bar{V}^0$, the $(z)$ and $(x)$ Maxwell's Equations give
\begin{align*}
\partial_z \partial_x a_x^{(0)}(x,z) = 0,\\
\partial^2_z a_x^{(0)}(x,z) = 0,
\end{align*}
\noindent which implies $\partial_z a_{x}^{(0)} = \text{constant}$. To fix this constant, we need to apply ingoing boundary conditions. Note that $u=t-z$ and $v=t+z$ are the two null coordinates in $AdS_4$. Therefore for the fluctuation to be ingoing, we require $\delta A_x(x,z,t) = \delta A_x (x,v)$ which fixes $\partial_z a_{x}^{(0)} = -E$. This gives the order $O(\bar{V}^0)$ contribution to the DC conductivity $\sigma = 1 +O(\bar{V}^2)$, which agrees with the pure $AdS_4$ value.

The order $O(\bar{V}^2)$, the $(z)$ and $(x)$ Maxwell's Equations read
\begin{align}
\label{conserveqs}
\partial_z \left[E(\alpha-\beta+\delta)-2z^2 h_{tx}^{(1)}\partial_z\varphi-2\partial_z a_x^{(2)} \right]=0, \\
\partial_x \left[E(\alpha-\beta+\delta)-2z^2 h_{tx}^{(1)}\partial_z\varphi-2\partial_z a_x^{(2)} \right]=0,
\end{align}
Note that these are exactly the equations for the conservation of the bulk current to $O(\bar{V}^2)$. They fix
\begin{align}
\label{ax2}
\partial_z a_x^{(2)} = c - 2z^2 h_{tx}^{(1)}\partial_z\varphi + \frac{E}{2}(\alpha-\beta+\delta),
\end{align}
for an arbitrary constant $c$. Note that the average of the above is exactly the numerator in \eqref{conductivity}. Since $\mathbb{E}[\alpha-\beta+\delta]=0$ everywhere in the bulk from Eqs. \eqref{alphadelta}, \eqref{beta} and $\lim\limits_{z\to 0}\mathbb{E}[z^2 h_{tx}^{(1)}\partial_z\varphi]=0$ to avoid deformations of the dual field theory Minkowski metric, $c$ is exactly the correction the the conductivity we are after. To fix $c$, we need to impose ingoing boundary conditions in the Poincar\'e horizon $z=\infty$, or in other words $\delta A_x (t,z,x) = \delta A_x(v,x)$ for $v=t+z$. This fixes $\lim\limits_{z\to\infty}\partial_z a_x^{(2)}(x,z)=0$, and we can formally write
\begin{align}
\label{fixingc}
c=\lim\limits_{z\to\infty} 2\mathbb{E}[z^2 h^{(1)}_{tx}\partial_z \varphi].
\end{align}

As we mentioned before, the Einstein's equations for $h_{tx}^{(1)}$ decouple from the background  
\begin{align*}
\partial_z\partial_x \left(z^2 h^{(1)}_{tx}\right) = z^2 E \partial_x\varphi, \\
\partial_z \left(z^{-2}\partial_z(z^2 h^{(1)}_{tx})\right)= E\partial_z\varphi,
\end{align*}
\noindent and can be readily solved by inserting the source \eqref{solvp} and integrating,
\begin{align*}
z^2 h^{(1)}_{tx}(x,z)=E\int \frac{\dd k}{2\pi}\frac{\mu_k}{k^3}2^{s+1}e^{-kz+ikx}\left(2+2kz+k^2z^2\right)+C(x),
\end{align*}
\noindent where $C(x)$ is a (random) integration constant. We suppose $C(x)$ admits a spectral representation with gaussian measure and write $C(x) = \int \frac{\dd k}{2\pi}e^{ikx}\mu_k c_k$ for a deterministic constant $c_k$. The boundary condition $\lim\limits_{z\to 0} z^2 h_{tx}=0$ then fixes $c_k = -2 k^{-3}$. Note that with this choice we have in particular $\lim\limits_{z\to 0}\mathbb{E}[z^2h^{(1)}_{tx}\partial_z\varphi]=0$ as claimed before. We can now explicitely fix $c$ by computing the average in Eq. \eqref{fixingc}
\begin{align*}
c=\left\{\begin{matrix}0 && \text{ for } s>1, \\ \frac{8\log{2}-5}{\pi} && \text{ for } s=1. \end{matrix}\right.
\end{align*}
Therefore for irrelevant disorder there are no corrections to the background conductivity to second order, $\sigma = 1 + O(\bar{V}^2)$, while for relevant disorder we have
\begin{align*}
\sigma = 1+\bar{V}^2 \gamma+O(\bar{V}^4),
\end{align*}
\noindent for $\gamma = \pi^{-1}(8\log{2}-5)>0$. This result is consistent with the previously discussed fact that for irrelevant deformations the background $AdS_4$ remains the IR fixed point of the system, while for marginal deformations the background geometry receives logarithmic corrections. Note that for $s<1$ the deformation is relevant. In this case $c$ diverges polynomially and perturbation theory breaks down. 

One might ask if the resummation carried out in the last sections alters the computation of the conductivity. This is not the case since as we argued before the metric fluctuations decouple from the background equations of motion. Resumming the background IR divergence for marginal deformations therefore does not change the conductivity, which is finite in the IR.

\section{Random chemical potential at finite temperature}
A natural generalization of the previous discussion is to include the effects of temperature. In practice, this is equivalent to imposing an AdS black brane boundary condition to the vacuum, around which we carry out a perturbative calculation. From the field theory perspective, we will be studying the perturbative effect of a random chemical potential in a quantum field theory at finite temperature. In practice, the presence of a horizon spoils the symmetry between the boundary coordinates $(t,x,y)$, which makes the calculations more involved. Following some previous ideas \cite{Hartnoll2015}, we will see that the problem can be analyzed in two opposite limits: high and low momenta modes. The high momenta modes will be exactly those that will contribute to the leading divergences of the metric components, therefore determining the emerging IR scaling. On the other hand, low momenta modes are constant along the bulk and will have the effect of renormalizing the temperature and charge of the black brane. We shall see that the initially uncharged black brane geometry develops an effective net charge proportional to the strength of the perturbation. Moreover to leading order in the disorder strength the thermal conductivity, but not the electrical conductivity, develops a Drude peak consistent with the breaking of translational symmetry by the random chemical potential. 

\subsection{Equations of Motion}

Consider again the action \eqref{action}. If $A = 0$, this action supports a finite temperature vacuum given by a $d+1=4$ AdS Schwarzschild black brane. Introducing a random chemical potential \eqref{bcvp} in the boundary can be seen as perturbation around this vacuum as long as $T\gg \bar{V}$. However, in order for disorder to be still relevant we need $k_0 = 1/a \gg T$. Therefore we are working with the hierarchy $k_0 \gg T \gg \bar{V}$. In analogy with the zero temperature case, we can set up a perturbative calculation around this background by looking at solutions of the system \eqref{eemaxwell} with the ansatz:
\begin{subequations}
\label{ansatz3}
\begin{align}
ds^2 &= \frac{1}{z^2}\left[-f(z)A(z)\dd t^2 + \frac{\dd z^2}{f(z)} + B(z)(\dd x^2 + \dd y^2) \right], \label{ansatz3a}\\
A &= a_t(z,\vec{x}) \dd t. \label{ansatz3b}
\end{align}
\end{subequations}
Following our previous discussion we are working directly with the averaged metric $A(z) = \mathbb{E}[A(z,\vec{x})]$, $B(z) = \mathbb{E}[B(z,\vec{x})]$ and with disorder in both boundary directions $(x,y)$, for which we can imposed isotropy. We also suppose that $f$ is a function of the holographic coordinate $z$ with a first order pole at a point $z_0$. It will be convenient to consider the rescaling $u = z/z_0$, such that $f(u=1) = 0$. As before, we set up our perturbation theory by letting
\begin{subequations}
\begin{align*}
A(u) &= 1 + \bar{V}^{2} \alpha(u) + O(\bar{V}^2), &B(u) &= 1 + \bar{V}^{2} \beta(u) + O(\bar{V}^2), \\ 
a_t(x,u) &= \bar{V} \varphi(x,u) + O(\bar{V}^3).
\end{align*}
\end{subequations}
The task is to solve the system \eqref{eom} together with the boundary conditions $\alpha(0) = \beta(0)$ and $\lim\limits_{u\to 0} \varphi(\vec{x}, u) = \int \frac{\dd[2]k}{2\pi} ~ e^{i\vec{k}\cdot\vec{x}} \mu_{\vec{k}}$. Further, we impose regularity and ingoing boundary conditions at the horizon $u=1$.

To order $\bar{V}^0$, the equations of motion are those for the AdS Schwarzschild background,
\begin{subequations}
\label{emblackeningf}
\begin{align}
-6+6f-4uf' + u^2f'' = 0, \\
3-3f+uf' = 0,
\end{align}
\end{subequations}
\noindent which are trivially satisfied by $f = 1-u^3$. To order $\bar{V}$, we have Maxwell's Equations for the vector potential, while no further metric equations are sourced:
\begin{equation*}
f\partial_u^2\varphi +z_0^2 \partial_x^2 \varphi = 0.
\end{equation*}
Again, we decompose $\varphi = \int \frac{\dd[2]{k}}{(2\pi)^2}~e^{i\vec{k}\cdot\vec{x}}\varphi_{\vec{k}}(u)$ to get:
\begin{equation}
\label{maxwellfinite}
f\varphi''_\kappa -\vec{\kappa}^2 \varphi_k = 0,
\end{equation}
\noindent where we have defined the dimensionless momentum $\kappa = z_0 |\vec{k}|$. Unfortunately we cannot solve the above equation explicitly. However we will be interested in two limits, the low (or zero) $\kappa\ll 1$ and high $\kappa\gg 1$ momentum modes. In the first limit, we have $\varphi'_{0} = \eta$ which is constant, while in the second limit $\kappa \gg 1$ we can rely on the WKB approximation
\begin{equation*}
\varphi_k(u) = \mu_k f^{-1/4} e^{-\kappa \int f^{-1/2}}.
\end{equation*}

To order $\bar{V}^2$, Einstein's Equations give:
\begin{align*}
f \alpha'' + \frac{(uf'-2f)}{2 u}(3\alpha'+\beta')&= -\frac{u^2 z_0^2}{2f} \mathbb{E}\left[f(\partial_u \varphi)^2+z_0^2(\nabla \varphi)^2\right], \\
f\alpha'' +2f\beta''+\frac{3uf'-2f}{2u} \alpha' + \frac{uf'-2f}{2u}\beta' &= \frac{u^2 z_0^2}{2f} \mathbb{E}\left[-f(\partial_u \varphi)^2+z_0^2(\nabla \varphi)^2\right], \\
f\beta'' - \frac{f}{u}\alpha' - \frac{uf'-4f}{u} &=  \frac{u^2 z_0^2}{2f} \mathbb{E}\left[f(\partial_u \varphi)^2-z_0^2(\partial_x\varphi)^2+z_0^2(\partial_y\varphi)^2\right],  \\
f\beta'' - \frac{f}{u}\alpha' - \frac{uf'-4f}{u} &=  \frac{u^2 z_0^2}{2f} \mathbb{E}\left[f(\partial_u \varphi)^2+z_0^2(\partial_x\varphi)^2-z_0^2(\partial_y\varphi)^2\right].
\end{align*}
\noindent where we made use of the zeroth order equations. These can be explicitly decoupled in two second order equations
\begin{subequations}
\begin{align}
(3+f)^2f^{-1/2}\partial_u\left(\frac{f^{3/2}}{u^2(3+f)}\partial_u\alpha\right) &=z_0^2~ \mathbb{E}[3(\partial_u\varphi)^2+z_0^2(\nabla \varphi)^2] \label{eomfinite1},\\
4 f^{3/2}\partial_u \left(\frac{f^{1/2}}{u^2}\partial_u \beta\right) &=-z_0^2~ \mathbb{E}[f(\partial_u\varphi)^2+z_0^2(\nabla \varphi)^2] \label{eomfinite2}.
\end{align}
\end{subequations}

\subsection{High momenta modes}

The effect of modes with $\kappa\gg 1$ was first discussed in Ref. \cite{Hartnoll2015} in the context of a random scalar deformation. Since the calculations for the high momenta modes for the charged deformation is similar, we only review the results and direct the reader to Ref. \cite{Hartnoll2015} for the technical details.

An explicit calculation shows that for $\kappa\gg1$ the main contribution to integrating equations \eqref{eomfinite1} and \eqref{eomfinite2} comes from the near boundary region $u=0$. Note that in this region these equations reduce to \eqref{eq32d} and \eqref{eq42d}, giving logarithmic corrections to the metric coefficient $\alpha \sim \log{z_0/a}$. The important remark is that the second order correction to the surface gravity of the background is proportional to $\alpha$. In particular, this implies that the temperature of the black hole receives second order logarithmic corrections from the high momenta modes. If we further assume that these corrections can be resummed as in the zero temperature setting, the temperature will develop a Lifshitz scaling  $T\sim z_0^{-\bar{z}}$ with the horizon. The upshot is that all other thermodynamic quantities are affected by the way they scale with temperature. It is important to note that this is a direct consequence of the logarithmic corrections for the metric coefficient $\alpha$. Since we find a similar correction, the results of Ref. \cite{Hartnoll2015} should apply here.

What about lower momenta modes? From Eq. \eqref{maxwellfinite}, it is clear that for $\kappa \ll 1$ the source is approximately constant in the bulk, and therefore does not contribute to the singular behaviour of the metric. From the RG point of view, these modes are irrelevant and can only possibly renormalize the background geometry. As we will discuss below, this is indeed the case.

\subsection{Low momenta modes}
\label{lowmomem}
We will show that low momenta modes play the role of renormalizing the background by introducing a charge $Q\sim \bar{V}$ in the originally neutral black brane. 

Consider the renormalized emblackening factor $f = \bar{f} + \bar{V}^2 \delta f$ where $\bar{f}(u) = 1-u^3$ with ansatz \eqref{ansatz3}. This shift has no effect in the zeroth and first order equations. However, it introduces an extra factor in Eq. \eqref{eomfinite2}:
\begin{align*}
\left(u^{-3}\delta f\right)'-\bar{f}^{1/2}\partial_u \left(\frac{\bar{f}^{1/2}}{u^2}\partial_u \beta\right) &=\frac{z_0^2}{4\bar{f}}~ \mathbb{E}[\bar{f}(\varphi_{\kappa}')^2+\kappa^2 \varphi_{\kappa}^2].
\end{align*}
In particular, for $\kappa\ll 1$ the left hand side is constant, since by Maxwell's Equations \eqref{maxwellfinite} $(\varphi'_{\kappa \ll 1})^2=\mu_0^2$. This is precisely the statement that low momenta modes are constant along the bulk. Close to the horizon $u=1$ the first term on the right hand side drops, giving:
\begin{equation*}
(u^{-3}\delta f)' = \frac{z_0^2}{4} \mu_0^2,
\end{equation*}
\noindent which can be easily solved by $\delta f(u) = \frac{z_0^2}{4}\mu_0^2 \left( u^4 - u^3 \right)$ and requiring $\delta f(0) = \delta f(1) = 0$. This correction gives precisely the emblackening factor $f(u) = 1 - (1+Q^2) u^3+ Q^2 u^4$ expected for an AdS Reissner-Nordstrom black brane with charge,
\begin{align*}
Q^2 =\frac{z_0^2}{4}\mu_0^2 \bar{V}^2.
\end{align*}
Therefore the constant low momenta modes have the effect of renormalizing the near horizon geometry of the initially uncharged black brane, adding a charge proportional to the sourced disorder. However note that this only contributes to the previous discussion at order $O(\bar{V}^4)$.This explains why to leading order it is justified to look only at high momenta modes when analyzing the divergences of the metric under the flow of the renormalization group. We expect the full non-linear solution to be a charged black brane with a temperature reflecting both contributions discussed above.

\subsection{Conductivity and momentum dissipation}
\label{condfinitetemp}
Recent works by Donos, Gauntlett \cite{Donos2014b, Donos2014c, Banks2015,Donos2015a}, built upon previous membrane paradigm ideas \cite{Iqbal2009}, have simplified enormously the task of computing averaged DC conductivities in inhomogeneous backgrounds at finite temperature. Specifically, in \cite{Donos2014b} they provide an explicit formula for the DC conductivity of the Einstein-Maxwell system sourced by a periodic potential in terms of near horizon data. The generalization of their results to our model read
\begin{align}
\label{conddonos}
\sigma = 1 + \bar{V}^2X^{-1} \mathbb{E}\left[\frac{\varphi_{(0)}}{A_{(0)}} \right]^2,
\end{align}
\noindent where $X =  \mathbb{E}\left[\left(\frac{\varphi^{(0)}}{A_{(0)}}\right)^2 \right]-\mathbb{E}[B_{(0)}^{-3}\partial_x B_{(0)}]-\mathbb{E}\left[\frac{\varphi_{(0)}}{A_{(0)}} \right]^2$ and the metric and gauge field are evaluated at the horizon $u=1$. In order to compute the corrections to the conductivity is necessary to take averages of fractions, which is usually a hard task. However we can still get a qualitative picture without having to compute the averages explicitly. First, it is clear that generically $X\neq0$ since the first term is an average over a second moment. The same is \emph{a priori} not clear for the numerator, which is an average over a first moment. 

In perturbation theory, $A = 1 + \bar{V}^2 \alpha$, and we can expand the denominator for small $\bar{V}$: $\mathbb{E}\left[\varphi_{(0)}/A_{(0)} \right] \sim \mathbb{E}\left[\varphi_{(0)}]-\bar{V}^2\mathbb{E}[ \alpha_{(0)}\varphi_{(0)} \right]+O(\bar{V}^4)$. By construction we have $\mathbb{E}[\varphi_{(0)}]=0$, and the problem simplifies to computing $\mathbb{E}[\alpha_{(0)}\phi_{(0)}]$. In principle to compute this average explicitly one needs the exact background to second order. However by looking at the most general spectral decomposition of $\alpha$ that solves the equations of motion one can compute the average \eqref{conddonos} in function of the coefficients $\alpha_{\vec{k}}$. Without loss of generality we can write $\alpha = \alpha_{hom}(u)+\alpha_{inh}(\vec{x},u)$. It is clear that only $\alpha_{inh}$ contributes to the perturbative corrections of the conductivity, since any homogeneous part (which is a constant at the horizon) vanishes when averaged with the source $\phi_{(0)}$. From the equations of motion we can write (c.f. appendix \ref{appendix1} for further details)
\begin{align*}
\alpha_{inh}(\vec{x},u) = \sum\limits_{\vec{k}} \alpha^{0}_{\vec{k}}(u)\prod\limits_{i}\cos{2\theta_{i,\vec{k}}} + \sum\limits_{\vec{k}\neq\vec{l}} \alpha^{+}_{\vec{k},\vec{l}}(u)\prod\limits_{i}\cos{\theta^+_{i,\vec{k},\vec{l}}}+\sum\limits_{\vec{k}\neq\vec{l}} \alpha^{-}_{\vec{k},\vec{l}}(u)\prod\limits_{i}\cos{\theta^-_{i,\vec{k},\vec{l}}},
\end{align*}
\noindent where we used a discrete representation for simplicity (c.f. Eq. \eqref{discrep}), and defined $\theta^{\pm}_{i,k_i,l_i} =\theta_{i,k_i}\pm\theta_{i,l_i}$. Letting $\varphi =  \sum\limits_{\vec{k}} \varphi_{\vec{k}}(u)\prod\limits_{i}\cos{\theta_{i,\vec{k}}}$ and evaluating at $u=1$, one check that $\mathbb{E}[\alpha_{inh(0)}\phi_{(0)}]=0$ and therefore
\begin{equation*}
\sigma=1+O(\bar{V}^4).  
\end{equation*}
One could be tempted to extend this argument to fourth or higher orders in $\bar{V}$. However this is a really hard task as it would also require the computation at least of the third order contribution to the vector potential as well as the fourth order contribution to the metric. 

It is intriguing that the random chemical potential does not contribute, to leading order at least, to the background electric conductivity. The likely physical reason for that behaviour is a peculiar feature of this realisation of disorder: charge carriers, whose average charge vanishes, and that naturally contributes to the electrical conductivity, are at the same time the source of disorder in the system. This dual role is rather unusual in condensed matter systems where scatterers are typically uncharged and quenched and therefore do not contribute to the electrical conductivity. 

We confirm that this unexpected result is a peculiarity of the electrical conductivity in this model of disorder by computing the thermal conductivity $\kappa$ 
 \footnote{Not to be mistaken with the dimensionless momentum we defined before.}, which describes transport of energy instead of charge. Following again the results of \cite{Donos2014b}, $\kappa$ is given by,
\begin{align*}
\kappa = \frac{(4\pi)^2 T}{X+ \mathbb{E}\left[\frac{\varphi_{(0)}}{A_{(0)}} \right]^2}.
\end{align*} 
It is straightforward to check that now the thermal conductivity depends on disorder, even to leading order, since we have an average over a second moment of the source $\varphi$ inside $X$, which gives a non-zero contribution to second order. We can estimate this in the high temperature limit $T\gg k_0$ where the main contribution to the geometry is $\varphi_{\kappa\ll 1} = (1-u)\mu(x,y)$. Therefore $\kappa = \frac{(4\pi)^4 T^3}{9} \frac{1}{\mathbb{E}[\mu^2]}$ which leads to,
\begin{align}
\label{thermo}
\kappa = \frac{(4\pi)^3}{9} \frac{T^3}{k_0 \bar{V}^2}.
\end{align}
As was expected, in the absence of disorder $\bar{V}\to 0$, $\kappa$ diverges as $1/\bar{V}^2$ since for no disorder translational invariance is recovered. The expression \eqref{thermo} also suggests that the relaxation scale of momentum is given by $\tau^{-1} \sim k_0 \bar{V}^2$. This is in full agreement with recent results in a set up similar to ours where disorder is introduced by a random scalar field in the boundary \cite{Hartnoll2015a}. 

Finally, it is important to stress that all these results are restricted to averaged conductivities. It would be interesting to know higher moments and the full probability distribution of the relevant observables. That for instance could provide additional information on the effect of a random chemical potential on the electrical conductivity for which we have clearly observed that a simple average misses important features.

\section{Conclusions}
We have studied analytically the role of weak disorder in Einstein-Maxwell theory and its relation, by holography,  with the transport properties of the dual field theory. Disorder is introduced through a random correlated chemical potential whose conformal dimension can be tuned by modifying the strength of the correlations. In that way we can investigate, within the Einstein-Maxwell theory, irrelevant, marginal or relevant 
perturbations. We have focused in the first two cases where we have found that, to leading order, irrelevant perturbations do not alter the conductivity while marginal perturbations induce a positive correction. Both results are in agreement with the recently proposed bound \cite{Lucas2014,Lucas2015,Lucas2015a} for the DC conductivity at finite temperature. Curiously disorder does not seem to suppress incoherent transport even at zero temperature. It would be interesting to understand why these field theory degrees of freedom are protected from disorder. In the marginal case at zero temperature we also found infrared logarithmic singularities in the metric that, after resummation as in Ref. \cite{Hartnoll2014}, lead to a Lifshitz-like geometry. At finite temperature we have shown that despite the fact that the chemical potential has zero average the black hole develops some net charge. The thermal conductivity is consistent with a disordered potential that induces relaxation of momentum. However the average  electrical conductivity, as in the zero temperature case, is still not affected by disorder to leading order in perturbation theory. It would be interesting to study the conditions to observe a transition from a neutral to a charged infrared background as disorder is increased in the limit of a chemical potential with zero average. Another interesting question is to clarify the conditions for a correction to the conductivity at finite temperature due to disorder. We plan to address these problems in the near future.   

\acknowledgments
A. M. G. thanks Hong Liu and Elias Kiritsis for illuminating discussions. 
A. M. G. acknowledges
partial support from EPSRC, grant No.
EP/I004637/1.
B. L. thanks Andrew Lucas for interesting discussions and suggestions concerning the conductivity. B.L. is supported by a CAPES/COT grant No. 11469/13-17. 
Both authors are grateful to the Galileo Galilei Institute
for Theoretical Physics for the hospitality and the INFN
for partial support during the completion of this work. 
\appendix

\section{Conductivity at finite temperature}  
\label{appendix1}
Consider the Einstein-Maxwell system at finite temperature with an inhomogenous chemical potential in both boundary directions. We work in coordinates such that the horizon is at $u=1$ and the boundary at $u=0$. To zeroth order in perturbation theory, Einstein's Equations fix the usual Schwarzschild emblackening factor as in \eqref{emblackeningf}. Looking for solutions of the type
\begin{align*}
\label{ansatz4}
ds^2 &= \frac{z_0^2}{u^2}\left[-f(u)(1+\bar{V}^2\alpha(u,\vec{x}))\dd t^2 + \frac{z_0^2 \dd z^2}{f(u)} + (1+\bar{V}^2\beta(u,\vec{x}))(\dd x^2 + \dd y^2) \right], \\
A &= \bar{V}\varphi(u,\vec{x}) \dd t.
\end{align*}
The second order equations read:
\begin{align*}
 2uf\partial_u^2\alpha + 2u z_0^2\nabla^2\alpha + (3uf'-6f)\partial_u\alpha+2(uf'-4f)\partial_u\beta &= -u^3z_0^2f^{-1} \left[f(\partial_u\varphi)^2+z_0^2(\nabla\varphi)^2\right]  \tag{tt},\\
2uf\partial_u^2\alpha+4uf\partial_u^2\beta+(3uf'-2f)\partial_u\alpha+2(uf'-2f)\partial_u\beta &=u^3z_0^2f^{-1}\left[(f\partial_u\varphi)^2-z_0^2(\nabla\varphi)^2\right] \tag{uu},\\ 
f'\partial_x\alpha+2f\partial_x\partial_u(\alpha+\beta) &= -\frac{1}{2}u^2 z_0^2 f^{-1}\partial_x\varphi\partial_u\varphi \tag{ux},\\
2uf\partial_u^2\beta+2uz_0^2\nabla^2\beta+2uz_0^2\partial_x^2\alpha -2f\partial_u\alpha+2(uf'-4f)\partial_u\beta &=u^3z_0^2f^{-1}\left[f(\partial_z\varphi)^2-z_0^2\left( (\partial_x\varphi)^2-(\partial_y\varphi)^2\right)\right] \tag{xx},\\ 
f'\partial_y\alpha+2f\partial_y\partial_u(\alpha+\beta) &= -\frac{1}{2}u^2 z_0^2 f^{-1}\partial_y\varphi\partial_u\varphi \tag{uy},\\ 
f\partial_y\partial_x\alpha &= u^2 z_0^2\partial_x\varphi\partial_y\varphi \tag{xy}, \\
2uf\partial_u^2\beta+2uz_0^2\nabla^2\beta+2uz_0^2\partial_y^2\alpha -2f\partial_u\alpha+2(uf'-4f)\partial_u\beta &=u^3z_0^2f^{-1}\left[f(\partial_z\varphi)^2+z_0^2\left( (\partial_x\varphi)^2-(\partial_y\varphi)^2\right)\right] \tag{yy},
\end{align*}
\noindent where we introduced $\nabla = (\partial_x, \partial_y)$. It is convenient to look at the following linear combinations,
\begin{align}
\label{eqsreduced}
4f^{1/2}\partial_u \left(u^{-2}f^{1/2}\partial_u\beta \right)+2u^{-2}\nabla^2\beta &=-z_0^2f^{-1}\left[f(\partial_u^2\varphi)^2+z_0^2(\nabla\varphi)^2\right], \\
(3+f)^2f^{-1/2}\partial_u\left(\frac{f^{3/2}}{u^2(3+f)}\partial_u\alpha\right)+2u^{-2}z_0^2f\nabla^2\alpha+u^{-2}z_0^2(f-3)\nabla^2\beta &=z_0^2f^{-1}~ \mathbb{E}[3f(\partial_u\varphi)^2+z_0^2(\nabla \varphi)^2],
\end{align}
\noindent where we took $-tt+uu+xx+yy$ and $3/2(f+1)+1/2(f-3)(uu-xx-yy)$ respectively. Note in particular that these equations reduce to \eqref{eomfinite1} and \eqref{eomfinite2} over averaging. The source can be expanded in the spectral basis as 
\begin{equation*}
\varphi(u,\vec{x}) =\sum\limits_{\vec{k}} \varphi_{\vec{k}}(u)\prod\limits_{i\in\{x,y\}} \cos{\theta_\vec{i,k}},
\end{equation*}
\noindent where $\vec{k} = (k_x,k_y) = (n_x,n_y)k_0/N$ with $n_x,n_y\in\{1,2,\dots,N-1\}$ and $\theta_{i,k_i} = k_i x^i + \gamma_i$ for $\gamma_i\in[0,2\pi)$ i.i.d. random variables (c.f. discussion in section \ref{iranduv}). We have opted for a discrete representation here for clarity, but this should not change the result. Recall that $\varphi_{\vec{k}}(u)$ can be obtained from Maxwell's Equations \eqref{maxwellfinite}, but an explicit solution is not needed for our purposes.

Further, we can write
\begin{align*}
(\partial_u\varphi)^2 &= \left(\sum\limits_{\vec{k}} \varphi'_{\vec{k}}\prod\limits_{i} \cos{\theta_{i,\vec{k}}}\right)  \left(\sum\limits_{\vec{l}} \varphi'_{\vec{l}}\prod\limits_{i} \cos{\theta_{i,\vec{l}}}\right) = \sum\limits_{\vec{k},\vec{l}} \varphi'_{\vec{k}}\varphi'_{\vec{l}}\prod\limits_{i} \cos{\theta_{i,\vec{k}}}\cos{\theta_{i,\vec{l}}} \\
&= \frac{1}{2}\sum\limits_{\vec{k},\vec{l}} \varphi'_{\vec{k}}\varphi'_{\vec{l}}\prod\limits_{i} \left(\cos{\theta^-_{i,\vec{k},\vec{l}}}+\cos{\theta^+_{i,\vec{k},\vec{l}}}\right)\\
&= \frac{1}{2}\sum\limits_{\vec{k}} (\varphi'_{\vec{k}})^2\prod\limits_{i} \left(1+\cos{2\theta_{i,\vec{k}}}\right)+ \frac{1}{2}\sum\limits_{\vec{k}\neq\vec{l}} \varphi'_{\vec{k}}\varphi'_{\vec{l}}\prod\limits_{i} \left(\cos{\theta^-_{i,\vec{k},\vec{l}}}+\cos{\theta^+_{i,\vec{k},\vec{l}}}\right),\\
(\nabla\varphi)^2 &= \frac{1}{2}\sum\limits_{\vec{k},\vec{l}} (\vec{k}\cdot\vec{l}) \varphi_{\vec{k}}\varphi_{\vec{l}}\prod\limits_{i} \left(\cos{\theta^-_{i,\vec{k},\vec{l}}}-\cos{\theta^+_{i,\vec{k},\vec{l}}}\right)\\
&=\frac{1}{2}\sum\limits_{\vec{k}} k^2(\varphi_{\vec{k}})^2\prod\limits_{i} \left(1-\cos{2\theta_{i,\vec{k}}}\right)+ \frac{1}{2}\sum\limits_{\vec{k}\neq\vec{l}}(\vec{k}\cdot\vec{l}) \varphi_{\vec{k}}\varphi_{\vec{l}}\prod\limits_{i} \left(\cos{\theta^-_{i,\vec{k},\vec{l}}}-\cos{\theta^+_{i,\vec{k},\vec{l}}}\right),
\end{align*}
\noindent where we have defined $\theta^{\pm}_{i,k_i,l_i} = \theta_{i,k_i}\pm\theta_{i,l_i}$. This determines the spectral decomposition of the metric coefficients in terms of the sources. For example, we can write $\alpha(u,\vec{x}) = \alpha_{hom}(u) + \alpha_{inh}(\vec{x},u)$ with
\begin{align}
\alpha_{inh}(\vec{x},u) = \sum\limits_{\vec{k}} \alpha^{0}_{\vec{k}}(u)\prod\limits_{i}\cos{2\theta_{i,\vec{k}}} + \sum\limits_{\vec{k}\neq\vec{l}} \alpha^{+}_{\vec{k},\vec{l}}(u)\prod\limits_{i}\cos{\theta^+_{i,\vec{k},\vec{l}}}+\sum\limits_{\vec{k}\neq\vec{l}} \alpha^{-}_{\vec{k},\vec{l}}(u)\prod\limits_{i}\cos{\theta^-_{i,\vec{k},\vec{l}}},
\end{align}
\noindent with a similar expression for $\beta$. By linearity, the task of solving equations \eqref{eqsreduced} now reduces to solving coupled ODEs for $\alpha_{hom}, \beta_{hom}$, $\alpha^{0},\beta^{0}$ and $\alpha^{\pm},\beta^{\pm}$. This is in principle doable but cumbersome, and does not bring any insight. Examples of explicit solutions for zero and finite temperature backgrounds in a similar context were given in references \cite{Hartnoll2014,OKeeffe2015,Hartnoll2015}. However for the purposes of applying the formula \eqref{conddonos} we do not need the full solution. 

By linearity of the mean, we just need to compute terms like $\mathbb{E}[\cos{\theta_k}]$,$\mathbb{E}[\cos{2\theta_n}\cos{\theta_k}]$ and $\mathbb{E}[\cos{\theta^{\pm}_{nm}}\cos{\theta_{k}}]$ for $n\neq m$. The first is trivially zero since it is the integral of one cosine over a full period. To compute the other terms, we use the angle sum rule $\cos{\theta^{\pm}_{nm}} = \cos{\theta_{n}}\cos{\theta_{m}}\mp  \sin{\theta_{n}} \sin{\theta_{m}}$. In order to have a nonzero integral we need all cosines and sines to group into a single power, since any single cosine vanishes when integrated over. For the second term, this will only happen when $k=n$, but in this case the integrals are over $\cos^3{\theta}$ and $\sin^2{\theta}\cos{\theta}$ which vanish on a period. In the third term, there will be always a cosine or sine left over since $n\neq m$. Thus  $\mathbb{E}[\cos{\alpha_k}]=\mathbb{E}[\cos{2\theta_n}\cos{\theta_k}] = \mathbb{E}[\cos{\theta^{\pm}_{nm}}\cos{\theta_{k}}] = 0$ generically. The result quoted in section \ref{condfinitetemp} follows.

\bibliography{bibliography}

\providecommand{\href}[2]{#2}\begingroup\raggedright\begin{thebibliography}{10}

\bibitem{Abrahams1979}
E.~Abrahams, P.~W. Anderson, D.~C. Licciardello, and T.~V. Ramakrishnan,
  ``{Scaling Theory of Localization: Absence of Quantum Diffusion in Two
  Dimensions},'' \href{http://dx.doi.org/10.1103/PhysRevLett.42.673}{{\em
  Physical Review Letters} {\bfseries 42} no.~10, (Mar., 1979) 673--676}.
  \url{http://link.aps.org/doi/10.1103/PhysRevLett.42.673}.

\bibitem{Anderson1958}
P.~W. Anderson, ``{Absence of diffusion in certain random lattices},''
  \href{http://dx.doi.org/10.1103/PhysRev.109.1492}{{\em Physical Review}
  {\bfseries 109} (1958) 1492--1505},
  \href{http://arxiv.org/abs/0807.2531}{{\ttfamily arXiv:0807.2531}}.

\bibitem{Frohlich1983}
J.~Fr\"{o}hlich and T.~Spencer, ``{Absence of diffusion in the Anderson tight
  binding model for large disorder or low energy},''
  \href{http://dx.doi.org/10.1007/BF01209475}{{\em Communications in
  Mathematical Physics} {\bfseries 88} no.~2, (1983) 151--184}.
  \url{http://dx.doi.org/10.1007/BF01209475}.

\bibitem{Vollhardt1980}
D.~Vollhardt and P.~Wolfle, ``Anderson localization in $d \leq 2$ dimensions: A
  self-consistent diagrammatic theory,''
  \href{http://dx.doi.org/10.1103/PhysRevLett.45.842}{{\em Phys. Rev. Lett.}
  {\bfseries 45} (Sep, 1980) 842--846}.
  \url{http://link.aps.org/doi/10.1103/PhysRevLett.45.842}.

\bibitem{Vollhardt1982}
D.~Vollhardt and P.~W\"olfle, ``Scaling equations from a self-consistent theory
  of anderson localization,''
  \href{http://dx.doi.org/10.1103/PhysRevLett.48.699}{{\em Phys. Rev. Lett.}
  {\bfseries 48} (Mar, 1982) 699--702}.
  \url{http://link.aps.org/doi/10.1103/PhysRevLett.48.699}.

\bibitem{Garcia2008}
A.~M. Garc\'{\i}a-Garc\'{\i}a, ``Semiclassical theory of the anderson
  transition,'' \href{http://dx.doi.org/10.1103/PhysRevLett.100.076404}{{\em
  Phys. Rev. Lett.} {\bfseries 100} (Feb, 2008) 076404}.
  \url{http://link.aps.org/doi/10.1103/PhysRevLett.100.076404}.

\bibitem{Rodriguez2011}
A.~Rodriguez, L.~J. Vasquez, K.~Slevin, and R.~A. R\"omer, ``Multifractal
  finite-size scaling and universality at the anderson transition,''
  \href{http://dx.doi.org/10.1103/PhysRevB.84.134209}{{\em Phys. Rev. B}
  {\bfseries 84} (Oct, 2011) 134209}.
  \url{http://link.aps.org/doi/10.1103/PhysRevB.84.134209}.

\bibitem{Slevin2014}
K.~Slevin and T.~Ohtsuki, ``Critical exponent for the anderson transition in
  the three-dimensional orthogonal universality class,'' {\em New Journal of
  Physics} {\bfseries 16} no.~1, (2014) 015012.
  \url{http://stacks.iop.org/1367-2630/16/i=1/a=015012}.

\bibitem{Garcia2007}
A.~M. Garc\'{\i}a-Garc\'{\i}a and E.~Cuevas, ``Dimensional dependence of the
  metal-insulator transition,''
  \href{http://dx.doi.org/10.1103/PhysRevB.75.174203}{{\em Phys. Rev. B}
  {\bfseries 75} (May, 2007) 174203}.
  \url{http://link.aps.org/doi/10.1103/PhysRevB.75.174203}.

\bibitem{Abou-Chacra1973}
R.~Abou-Chacra, D.~J. Thouless, and P.~W. Anderson, ``{A selfconsistent theory
  of localization},'' May, 1973.
\newblock \url{http://beta.iopscience.iop.org/0022-3719/6/10/009}.

\bibitem{Roati2008}
G.~Roati, C.~D'Errico, L.~Fallani, M.~Fattori, C.~Fort, M.~Zaccanti,
  G.~Modugno, M.~Modugno, and M.~Inguscio, ``Anderson localization of a
  non-interacting bose--einstein condensate,'' {\em Nature} {\bfseries 453}
  no.~7197, (2008) 895--898.

\bibitem{Billy2008}
J.~Billy, V.~Josse, Z.~Zuo, A.~Bernard, B.~Hambrecht, P.~Lugan, D.~Cl{\'e}ment,
  L.~Sanchez-Palencia, P.~Bouyer, and A.~Aspect, ``Direct observation of
  anderson localization of matter waves in a controlled disorder,'' {\em
  Nature} {\bfseries 453} no.~7197, (2008) 891--894.

\bibitem{Gorkov1979}
K.~D. Gor'kov~L.P., Larkin~A.I., ``Particle conductivity in a two-dimensional
  random potential,'' {\em JETP Lett.} {\bfseries 22} (Dec, 1979) 5142--5153.
  \url{http://www.jetpletters.ac.ru/ps/1364/article_20629.shtml}.

\bibitem{Altshuler1980}
B.~L. Altshuler, D.~Khmel'nitzkii, A.~I. Larkin, and P.~A. Lee,
  ``Magnetoresistance and hall effect in a disordered two-dimensional electron
  gas,'' \href{http://dx.doi.org/10.1103/PhysRevB.22.5142}{{\em Phys. Rev. B}
  {\bfseries 30} (Dec, 1980) 228--232}.
  \url{http://link.aps.org/doi/10.1103/PhysRevB.22.5142}.

\bibitem{Basko2006a}
D.~Basko, I.~Aleiner, and B.~Altshuler, ``{Metal--insulator transition in a
  weakly interacting many-electron system with localized single-particle
  states},'' \href{http://dx.doi.org/10.1016/j.aop.2005.11.014}{{\em Annals of
  Physics} {\bfseries 321} no.~5, (May, 2006) 1126--1205},
  \href{http://arxiv.org/abs/0506617}{{\ttfamily arXiv:0506617 [cond-mat]}}.
  \url{http://arxiv.org/abs/cond-mat/0506617}.

\bibitem{Basko2006}
D.~M. Basko, I.~L. Aleiner, and B.~L. Altshuler, ``{On the problem of many-body
  localization},'' \href{http://arxiv.org/abs/0602510}{{\ttfamily arXiv:0602510
  [cond-mat]}}. \url{http://arxiv.org/abs/cond-mat/0602510}.

\bibitem{Wang2008}
W.-M. Wang and Z.~Zhang, ``{Long Time Anderson Localization for the Nonlinear
  Random Schr\"{o}dinger Equation},''
  \href{http://dx.doi.org/10.1007/s10955-008-9649-1}{{\em Journal of
  Statistical Physics} {\bfseries 134} no.~5-6, (Nov., 2008) 953--968}.
  \url{http://link.springer.com/10.1007/s10955-008-9649-1}.

\bibitem{Bardarson2012}
J.~H. Bardarson, F.~Pollmann, and J.~E. Moore, ``{Unbounded Growth of
  Entanglement in Models of Many-Body Localization},''
  \href{http://dx.doi.org/10.1103/PhysRevLett.109.017202}{{\em Physical Review
  Letters} {\bfseries 109} no.~1, (July, 2012) 017202}.
  \url{http://link.aps.org/doi/10.1103/PhysRevLett.109.017202}.

\bibitem{Gopalakrishnan2015}
S.~Gopalakrishnan, M.~M\"uller, V.~Khemani, M.~Knap, E.~Demler, and D.~A. Huse,
  ``Low-frequency conductivity in many-body localized systems,''
  \href{http://dx.doi.org/10.1103/PhysRevB.92.104202}{{\em Phys. Rev. B}
  {\bfseries 92} (Sep, 2015) 104202}.
  \url{http://link.aps.org/doi/10.1103/PhysRevB.92.104202}.

\bibitem{Maldacena1997}
J.~M. Maldacena, ``{The Large N Limit of Superconformal Field Theories and
  Supergravity},'' \href{http://arxiv.org/abs/9711200}{{\ttfamily arXiv:9711200
  [hep-th]}}. \url{http://arxiv.org/abs/hep-th/9711200v3}.

\bibitem{Hartnoll2008b}
S.~A. Hartnoll and C.~P. Herzog, ``{Impure AdS/CFT correspondence},''
  \href{http://dx.doi.org/10.1103/PhysRevD.77.106009}{{\em Physical Review D}
  {\bfseries 77} no.~10, (May, 2008) 106009},
  \href{http://arxiv.org/abs/0801.1693}{{\ttfamily arXiv:0801.1693}}.
  \url{http://arxiv.org/abs/0801.1693}.

\bibitem{Fujita2008}
M.~Fujita, Y.~Hikida, S.~Ryu, and T.~Takayanagi, ``{Disordered systems and the
  replica method in AdS/CFT},''
  \href{http://dx.doi.org/10.1088/1126-6708/2008/12/065}{{\em Journal of High
  Energy Physics} {\bfseries 2008} no.~12, (Dec., 2008) 065--065},
  \href{http://arxiv.org/abs/0810.5394}{{\ttfamily arXiv:0810.5394}}.
  \url{http://arxiv.org/abs/0810.5394}.

\bibitem{Hartnoll2012}
S.~A. Hartnoll and D.~M. Hofman, ``{Locally Critical Resistivities from Umklapp
  Scattering},'' \href{http://dx.doi.org/10.1103/PhysRevLett.108.241601}{{\em
  Physical Review Letters} {\bfseries 108} no.~24, (June, 2012) 241601},
  \href{http://arxiv.org/abs/1201.3917}{{\ttfamily arXiv:1201.3917}}.
  \url{http://arxiv.org/abs/1201.3917}.

\bibitem{Arean2014}
D.~Are\'{a}n, A.~Farahi, L.~A. {Pando Zayas}, I.~S. Landea, and A.~Scardicchio,
  ``{Holographic superconductor with disorder},''
  \href{http://dx.doi.org/10.1103/PhysRevD.89.106003}{{\em Physical Review D}
  {\bfseries 89} no.~10, (May, 2014) 106003},
  \href{http://arxiv.org/abs/1308.1920}{{\ttfamily arXiv:1308.1920}}.
  \url{http://arxiv.org/abs/1308.1920}.

\bibitem{Zeng2013}
H.~B. Zeng, ``{Possible Anderson localization in a holographic
  superconductor},'' \href{http://dx.doi.org/10.1103/PhysRevD.88.126004}{{\em
  Physical Review D} {\bfseries 88} no.~12, (Dec., 2013) 126004},
  \href{http://arxiv.org/abs/1310.5753}{{\ttfamily arXiv:1310.5753}}.
  \url{http://arxiv.org/abs/1310.5753}.

\bibitem{Lucas2014}
A.~Lucas, S.~Sachdev, and K.~Schalm, ``{Scale-invariant hyperscaling-violating
  holographic theories and the resistivity of strange metals with random-field
  disorder},'' \href{http://dx.doi.org/10.1103/PhysRevD.89.066018}{{\em
  Physical Review D} {\bfseries 89} no.~6, (Mar., 2014) 066018},
  \href{http://arxiv.org/abs/1401.7993}{{\ttfamily arXiv:1401.7993}}.
  \url{http://arxiv.org/abs/1401.7993}.

\bibitem{Lucas2015}
A.~Lucas and S.~Sachdev, ``{Conductivity of weakly disordered strange metals:
  From conformal to hyperscaling-violating regimes},''
  \href{http://dx.doi.org/10.1016/j.nuclphysb.2015.01.017}{{\em Nuclear Physics
  B} {\bfseries 892} (Mar., 2015) 239--268},
  \href{http://arxiv.org/abs/1411.3331}{{\ttfamily arXiv:1411.3331}}.
  \url{http://arxiv.org/abs/1411.3331}.

\bibitem{Lucas2015a}
A.~Lucas, ``{Conductivity of a strange metal: from holography to memory
  functions},'' \href{http://arxiv.org/abs/1501.05656}{{\ttfamily
  arXiv:1501.05656}}. \url{http://arxiv.org/abs/1501.05656}.

\bibitem{Adams2011}
A.~Adams and S.~Yaida, ``{Disordered Holographic Systems I: Functional
  Renormalization},'' \href{http://arxiv.org/abs/1102.2892}{{\ttfamily
  arXiv:1102.2892}}. \url{http://arxiv.org/abs/1102.2892}.

\bibitem{Adams2014}
A.~Adams and S.~Yaida, ``{Disordered holographic systems: Marginal relevance of
  imperfection},'' \href{http://dx.doi.org/10.1103/PhysRevD.90.046007}{{\em
  Physical Review D} {\bfseries 90} no.~4, (Aug., 2014) 046007},
  \href{http://arxiv.org/abs/1201.6366}{{\ttfamily arXiv:1201.6366}}.
  \url{http://arxiv.org/abs/1201.6366}.

\bibitem{Hartnoll2014}
S.~A. Hartnoll and J.~E. Santos, ``{Disordered Horizons: Holography of Randomly
  Disordered Fixed Points},''
  \href{http://dx.doi.org/10.1103/PhysRevLett.112.231601}{{\em Physical Review
  Letters} {\bfseries 112} no.~23, (June, 2014) 231601},
  \href{http://arxiv.org/abs/1402.0872}{{\ttfamily arXiv:1402.0872}}.
  \url{http://arxiv.org/abs/1402.0872}.

\bibitem{Hartnoll2015}
S.~A. Hartnoll, D.~M. Ramirez, and J.~E. Santos, ``{Emergent scale invariance
  of disordered horizons},'' \href{http://arxiv.org/abs/1504.03324}{{\ttfamily
  arXiv:1504.03324}}. \url{http://arxiv.org/abs/1504.03324}.

\bibitem{Hartnoll2015a}
S.~A. Hartnoll, D.~M. Ramirez, and J.~E. Santos, ``{Thermal conductivity at a
  disordered quantum critical point},''
  \href{http://arxiv.org/abs/1508.04435}{{\ttfamily arXiv:1508.04435}}.
  \url{http://arxiv.org/abs/1508.04435}.

\bibitem{Donos2014b}
A.~Donos and J.~P. Gauntlett, ``{The thermoelectric properties of inhomogeneous
  holographic lattices},'' \href{http://arxiv.org/abs/1409.6875}{{\ttfamily
  arXiv:1409.6875}}. \url{http://arxiv.org/abs/1409.6875}.

\bibitem{Donos2015}
A.~Donos and J.~P. Gauntlett, ``{Navier-Stokes on Black Hole Horizons and DC
  Thermoelectric Conductivity},''
  \href{http://arxiv.org/abs/1506.01360}{{\ttfamily arXiv:1506.01360}}.
  \url{http://arxiv.org/abs/1506.01360}.

\bibitem{OKeeffe2015}
D.~K. O'Keeffe and A.~W. Peet, ``{Perturbatively charged holographic
  disorder},'' \href{http://arxiv.org/abs/1504.03288}{{\ttfamily
  arXiv:1504.03288}}. \url{http://arxiv.org/abs/1504.03288}.

\bibitem{Note1}
When working in higher dimensions, we denote $k=|\protect \mathbf {k}|$. In one
  dimension we explicitely include the modulus to avoid ambiguities.

\bibitem{Arean2014a}
D.~Arean, A.~Farahi, L.~A.~P. Zayas, I.~S. Landea, and A.~Scardicchio,
  ``{Holographic p-wave Superconductor with Disorder},''
  \href{http://arxiv.org/abs/1407.7526}{{\ttfamily arXiv:1407.7526}}.
  \url{http://arxiv.org/abs/1407.7526}.

\bibitem{Arean2015}
D.~Arean, L.~A.~P. Zayas, I.~S. Landea, and A.~Scardicchio, ``{The Holographic
  Disorder-Driven Supeconductor-Metal Transition},''
  \href{http://arxiv.org/abs/1507.02280}{{\ttfamily arXiv:1507.02280}}.
  \url{http://arxiv.org/abs/1507.02280}.

\bibitem{Note2}
We are not interested in the range $s<0$ since we already know disorder is
  relevant in this case a perturbative approach is not adequate.

\bibitem{Cardy1996}
J.~L. Cardy, {\em Scaling and renormalization in statistical physics}.
\newblock Cambridge University Press, Cambridge, 1996.
\newblock \url{http://www.loc.gov/catdir/description/cam027/95049981.html}.

\bibitem{Note3}
To avoid charging the notation, we conveniently denote $\protect \mathbb
  {E}[\alpha (x,z)] = \alpha (z)$, etc.

\bibitem{Kramer1993}
B.~Kramer and A.~MacKinnon, ``{Localization: theory and experiment},''
  \href{http://dx.doi.org/10.1088/0034-4885/56/12/001}{{\em Reports on Progress
  in Physics} {\bfseries 56} no.~12, (Dec., 1993) 1469--1564}.
  \url{http://beta.iopscience.iop.org/0034-4885/56/12/001}.

\bibitem{Anderson1980}
P.~W. Anderson, D.~J. Thouless, E.~Abrahams, and D.~S. Fisher, ``{New method
  for a scaling theory of localization},''
  \href{http://dx.doi.org/10.1103/PhysRevB.22.3519}{{\em Physical Review B}
  {\bfseries 22} no.~8, (Oct., 1980) 3519--3526}.
  \url{http://link.aps.org/doi/10.1103/PhysRevB.22.3519}.

\bibitem{Davison2015a}
R.~A. Davison and B.~Gout{\'{e}}raux, ``{Dissecting holographic
  conductivities},'' \href{http://dx.doi.org/10.1007/JHEP09(2015)090}{{\em
  Journal of High Energy Physics} {\bfseries 2015} no.~9, (Sep, 2015) 90},
  \href{http://arxiv.org/abs/1505.05092}{{\ttfamily arXiv:1505.05092}}.
  \url{http://arxiv.org/abs/1505.05092}.

\bibitem{Policastro2002}
G.~Policastro, D.~T. Son, and A.~O. Starinets, ``{From AdS/CFT correspondence
  to hydrodynamics},''
  \href{http://dx.doi.org/10.1088/1126-6708/2002/09/043}{{\em Journal of High
  Energy Physics} {\bfseries 2002} no.~09, (Sept., 2002) 043--043},
  \href{http://arxiv.org/abs/0205052}{{\ttfamily arXiv:0205052 [hep-th]}}.
  \url{http://arxiv.org/abs/hep-th/0205052}.

\bibitem{Iqbal2009}
N.~Iqbal and H.~Liu, ``{Universality of the hydrodynamic limit in AdS/CFT and
  the membrane paradigm},''
  \href{http://dx.doi.org/10.1103/PhysRevD.79.025023}{{\em Physical Review D}
  {\bfseries 79} no.~2, (Jan., 2009) 025023},
  \href{http://arxiv.org/abs/0809.3808}{{\ttfamily arXiv:0809.3808}}.
  \url{http://arxiv.org/abs/0809.3808}.

\bibitem{Note4}
We are grateful to Andrew Lucas for pointing this out.

\bibitem{Donos2014c}
A.~Donos and J.~P. Gauntlett, ``{Thermoelectric DC conductivities from black
  hole horizons},'' \href{http://dx.doi.org/10.1007/JHEP11(2014)081}{{\em
  Journal of High Energy Physics} {\bfseries 2014} no.~11, (Nov., 2014) 81},
  \href{http://arxiv.org/abs/1406.4742}{{\ttfamily arXiv:1406.4742}}.
  \url{http://arxiv.org/abs/1406.4742}.

\bibitem{Banks2015}
E.~Banks, A.~Donos, and J.~P. Gauntlett, ``{Thermoelectric DC conductivities
  and Stokes flows on black hole horizons},''
  \href{http://arxiv.org/abs/1507.00234}{{\ttfamily arXiv:1507.00234}}.
  \url{http://arxiv.org/abs/1507.00234}.

\bibitem{Donos2015a}
A.~Donos, J.~P. Gauntlett, T.~Griffin, and L.~Melgar, ``{DC Conductivity of
  Magnetised Holographic Matter},''
  \href{http://arxiv.org/abs/1511.00713}{{\ttfamily arXiv:1511.00713}}.
  \url{http://arxiv.org/abs/1511.00713}.

\bibitem{Note5}
Not to be mistaken with the dimensionless momentum we defined before.

\end{thebibliography}\endgroup
\bibliographystyle{utphys}
\end{document}